\documentstyle[graphicx,aps,amsmath]{revtex}

\newcommand{\bx}{{\mathbf{x}}}

\newcommand{\bk}{{\mathbf{k}}}
\newcommand{\bR}{{\mathbf{R}}}
\newcommand{\bK}{{\mathbf{K}}}
\newcommand{\bv}{{\mathbf{v}}}
\newcommand{\bdel}{{\mathbf{\nabla}}}

\newcommand{\tilGam}{\tilde{\Gamma}}
\newcommand{\barmu}{\bar{\mu}}
\newcommand{\barkap}{\bar{\kappa}}
\newcommand{\barC}{\bar{C}}
\newcommand{\barLam}{\bar{\Lambda}}
\newcommand{\carmu}{\hat{\mu}}
\newcommand{\carkap}{\hat{\kappa}}
\newcommand{\carC}{\hat{C}}
\newcommand{\carLam}{\hat{\Lambda}}
\newcommand{\carOmega}{\hat{\Omega}_0}
\newcommand{\carm}{\hat{m}}
\newcommand{\carE}{\hat{E}}

\newcommand{\fnbk}{f_{n\bk_0}}
\newcommand{\fubk}{f_{u\bk}}
\newcommand{\flbk}{f_{l\bk}}

\newcommand{\hatbp}{\hat{\mathbf{p}}}

\newcommand{\parpar}[2]{\frac{\partial #1}{\partial #2}}

  \newcommand{\ket}[1]{\left|#1\right\rangle}
\newcommand{\bra}[1]{\left\langle #1\right|}
\newcommand{\braket}[2]{\left \langle #1 \right|\left. \! #2 \right
  \rangle} \newcommand{\expect}[1]{\left\langle #1\right\rangle}

\newcommand{\fullexpect}[3]{\left\langle #1 \left| #2 \right| #3
  \right\rangle}

\begin{document}
\title{Bloch function description of a Bose-Einstein condensate in a
  finite optical lattice} \author{M. J. Steel and Weiping Zhang}
                      
\address{School of Mathematics, Physics, Computing and Electronics,
  Macquarie University, North Ryde, New South Wales 2109, Australia.}

\date{\today} \maketitle

\begin{abstract}
  We consider stationary and propagating solutions for a Bose-Einstein
  condensate in a periodic optical potential with an additional
  confining optical or magnetic potential. Using an effective mass
  approximation we express the condensate wavefunction in terms of
  slowly-varying envelopes modulating the Bloch modes of the optical
  lattice.  In the limit of a weak nonlinearity, we derive a nonlinear
  Schr\"{o}dinger equation for propagation of the envelope function
  which does not contain the rapid oscillation of the lattice.  We
  then consider the ground state solutions in detail in the regime of
  weak, moderate and strong nonlinear interactions.  We describe the
  form of solution which is appropriate in each regime, and place
  careful limits on the validity of each type of solution.

\end{abstract}

\section{Introduction}
There have now been a large number of experiments and an enormous
quantity of theory devoted to the properties of the trapped
Bose-Einstein condensates (BEC)~\cite{newbecs}. Motivated by the
existing experiments, the bulk of the theoretical work so far assumes
a harmonic trap in one, two or three dimensions. Lately, a few papers
have considered the consequences of exposing a BEC to an optical
periodic potential~\cite{sor98,zob98,jak98b}.  While not yet observed
experimentally, such a situation could easily be produced by applying
counter-propagating optical beams along one or more axes to produce a
standing-wave optical lattice. Indeed, the directional output coupler
of the BEC demonstrated by Leng \emph{et al.}~\cite{nistlaser} uses
such counter-propagating beams to induce Raman transitions into
untrapped states.  In this paper, we consider the special case of a
BEC confined in a \emph{finite} optical lattice, and develop theory to
describe the properties of the lattice BEC.

Subjecting a BEC to a periodic potential opens the way for a number of
new phenomena. S{\o}renberg and M{\o}lmer~\cite{sor98} assumed an
infinite lattice potential and showed as one would expect, that the
quasi-particle excitation spectrum exhibits a band-structure.  Zobay
\emph{et al.}~\cite{zob98} considered a propagating field of three
internal levels connected by Raman transitions. They showed that if
the center wavenumber is tuned close to the band gap induced by the
optical lattice, one can derive a pair of coupled mode
equations~\cite{win79} for forward and backward propagating waves.
These equations are well-known to support the interesting class of
solitary waves known as ``gap solitons''~\cite{ace89,des94}.  Gap
solitons have been observed when a periodic structure in an optical
fiber is subjected to high intensity light~\cite{egg96} but have not
been observed in an atom optics context. Finally,
Jaksch~\emph{et al.}~\cite{jak98b} have demonstrated an equivalence
between the system of a condensate trapped in an optical lattice and
the Bose-Hubbard model of condensed-matter physics.  At very low
site-occupation numbers, they predict the occurrence of phase
transitions between a Mott insulating state and coherent superfluid
flow.  Jaksch~\emph{et al.} also considered a complication not present
in the other two papers---the inclusion of a weak harmonic trap in
addition to the rapidly oscillating lattice.

It is this complication that we pursue in the present paper. As the
eigenstates or ``Bloch functions'' of a pure lattice potential are of
infinite extent, any real system must always include some form of
additional confining potential. (In any case, a real optical lattice
is always finite).  Thus we consider the case of an infinite periodic
potential combined with a weak confining potential.  While we will
allow the form of this potential to be rather general, in many cases a
harmonic oscillator form may be appropriate.  For example, one might
use optical beams to generate the lattice and provide the ultimate
confinement with a weak magnetic trap in the standard way.  In fact,
red-detuned counter-propagating focussed Gaussian beams would
effectively create both a lattice and a harmonic-like axial trap due
to the reduced intensity away from the beam focus.  In this case,
however, the lattice potential would also be modulated by a Gaussian
envelope. We shall restrict to the case where the periodic part of the
potential is uniform.

Our aim is to describe in detail the properties of the condensate wave
function in such a potential in both stationary and propagating
regimes.  We assume the condensate may be described by the standard
Gross-Pitaevskii equation (GPE) with the addition of a periodic
potential.  Numerically at least, finding the exact ground-state
wavefunction is then a straight-forward application of one's favorite
method---imaginary time propagation, shooting or relaxation.
Similarly, propagation problems may be solved by an appropriate
split-step algorithm for example. However, these procedures are
computationally very intensive as compared to the analogous problems
in the absence of the lattice for the following reason.  Typically, we
are interested in cases for which the period or ``lattice constant''
is relatively short compared to the extent of the wave function when
just the trapping potential is present. (If that is not the case, one
is really not dealing with a lattice at all, but a small set of
coupled non-identical wells.) Thus we are concerned with the case of a
rapidly varying periodic potential superposed on the slowly-varying
background of the trap potential. Hence as the wave function samples
many wells of the potential, it possesses a detailed structure on the
same scale as the lattice and a large number of grid points is
necessary for accurate calculations. A time-dependent propagation of
the field in a two-dimensional potential say, represents a formidable
calculation.  While such calculations are certainly
possible~\cite{zob98}, it is desirable to find a method that would
circumvent the computational burden as well as highlight the
underlying physics of the system.

Consider first the stationary problem, where it is clear what the
general form of the ground-state solution must be: to minimize the
energy, the atoms pile up in the lattice wells, and the relative
density in each well is determined by the local value of the confining
part of the potential.  The natural way to treat such a problem is to
capture the rapid variation on the scale of the lattice using the
Bloch functions of the periodic potential.  If the nonlinearity is not
too strong, the solution may be represented as a sum of a small number
of slowly-varying envelope functions modulating appropriate Bloch
functions.  These envelope functions obey simpler equations from which
all the rapid variation has been removed.  For propagating problems,
one may proceed in the same fashion but using different Bloch
functions.  In the language of solid state physics, this approach is
essentially an effective mass representation with the addition of the
nonlinear perturbation of the atomic collisions.  We should remark
that the work of Zobay \emph{et al.}~\cite{zob98} also uses an
envelope function approach, but their work does not include a
confining potential and is restricted to solutions close to a band
edge. In addition, they use plane waves rather than Bloch functions as
a basis, which limits the treatment to only relatively weak lattices.

The paper is structured as follows.  In section~\ref{sec:method} we
derive simple approximate equations for stationary and propagating
BECs in a finite optical lattice.  In
section~\ref{sec:Station-solutions} we present solutions for the
stationary ground state, comparing exact numerical solutions with our
Bloch function approach.  Section~\ref{sec:two-modes} discusses
extensions to treat fields near a band gap before we conclude.

\section{BEC in a finite optical lattice} \label{sec:method}
We consider a BEC in an infinite optical lattice with an additional
confining potential. To avoid incoherent heating from spontaneous
emission, the laser beams creating the lattice are detuned far from
the atomic resonance. In this case, the atomic field for the internal
excited state may be adiabatically eliminated and the optical lattice
acts as an external potential for the ground state field
$\hat{\psi}_g(\bx)$~\cite{cohen2,mar98}.  Then the macroscopic
wavefunction $\psi(\bx)=\expect{\hat{\psi}_g(\bx)}$ satisfies the GPE
\begin{equation}
  \label{eq:gpe}
  i\hbar \parpar{\psi(\bx,t)}{t}=
 \left[-\frac{\hbar^2}{2m}\nabla^2+U(\bx)+V(\bx) + 
        \Gamma|\psi(\bx,t)|^2 \right] \psi(\bx,t),
\end{equation}
where the optical dipole potential $U(\bx)$ is periodic in one or more
dimensions.  For simplicity we assume that the beams are aligned along
the coordinate axes and have wave vectors $\tilde{k}_x$, $\tilde{k}_y$
and $\tilde{k}_z$.  Thus dropping an unimportant constant in the
potential we may write
\begin{equation}
\label{eq:pot3d}
U(\bx) = - \sum_{\mu} \kappa_\mu \cos(k_\mu x_\mu) 
\end{equation}
where $x_\mu$ runs over the coordinates $x,y, $ and $z$, and $k_{\mu}
= 2 \tilde{k}_{\mu}=2\pi/d_\mu$, where $d_\mu$ is the period of the
lattice along axis $\mu$.  Also, $V(\bx)$ is the magnetic trap
potential and $\Gamma=4\pi\hbar^2 a/m$ is the interatomic interaction
strength, with $a$ the s-wave scattering length.

We now seek solutions to Eq.~\eqref{eq:gpe} using an expansion in
terms of suitable Bloch functions of the optical potential.  It will
be helpful to begin by stating a few results.

\subsection{Band structure and Bloch functions}
We consider stationary solutions to the linear Schr\"{o}dinger
equation
\begin{equation}
  \label{eq:periodic}
  \left [ -\frac{\hbar^2}{2m}\bdel^2 +U(\bx) \right] \phi(\bx)=E\phi(\bx).
\end{equation}

Bloch's theorem~\cite{callaway} states that the solutions take the
form
\begin{equation}
  \label{eq:blochfun}
\phi(\bx)=  \phi_{n,\bk}(\bx) = e^{i\bk\cdot\bx}u_{n\bk}(\bx),
\end{equation}
where the functions $u_{n\bk}(\bx)=u_{n\bk}(\bx+\bR_s)$ share the
periodicity of the lattice potential $U$, \emph{ie.,} $\bR_s$ is any
``lattice vector'', translation by which maps the potential onto
itself.  The energy eigenvalues $E= E_{n\bk}$ form a series of bands
indexed by the quantum number $n$, with an infinite number of bands
for each value of the wavevector $\bk$.  It is also shown in any
solid-state physics textbook, that the wavevectors are unique only up
to the addition of an arbitrary ``reciprocal lattice vector''.  One
can then restrict consideration to wavevectors lying within the
``first Brillioun zone''~\cite{callaway}.  In the one-dimensional
case, this corresponds to the range $|k|<\pi/d$.  We shall use a Dirac
notation in which the Bloch function $ \phi_{n\bk}(\bx)$ is denoted by
$\ket{n\bk}$.  In the usual way we restrict the wavevector to $N$
discrete values with $N$ large, so that the Bloch functions are
normalized as
\begin{equation}
\braket{m\bk}{n\bk'} \equiv \int_{\text{N cells}} \hspace{-.7cm}
        d^3\bx\, \phi^*_{m\bk}(\bx)\phi_{n\bk'}(\bx) 
        =N \Omega \delta_{mn}\delta_{\bk\bk'},
\end{equation}
where the integral is taken over $N$ unit cells of volume $\Omega$.

For the potential in Eq.~\eqref{eq:pot3d}, Eq.~\eqref{eq:periodic} is
separable and the well-known solutions are given by Mathieu
functions~\cite{abramowitz}.  In practice, the solutions and
eigenvalues are probably found most simply by solving
Eq.~\eqref{eq:periodic} directly.

\subsection{Equations of motion}
\subsubsection{Bloch function expansion}
Using the properties of the Bloch functions, we now develop an
envelope function approach for an atomic field $\psi(\bx,t=0)$ in a
finite lattice.  Certainly any wave function $\psi(\bx,t)$ may be
expanded in terms of the complete Bloch modes as
\begin{equation}
\label{eq:bloch-expand}
\psi(\bx,t) = \sum_n \sum_\bk A_{n\bk}(t) \ket{n\bk} e^{-iE_{n\bk}t/\hbar}.
\end{equation}
One could easily write down evolution equations for the amplitudes
$A_{n\bk}$ but such a procedure is of little use for finite systems in
which the wave function does not have the infinite extent of the Bloch
modes.  A very large number of the $A_{n\bk}$ would be required to
accurately describe the dynamics.  Instead it is more natural to use
the Bloch functions to capture the rapid oscillations in the wave
function but describe the localization of the field through slowly
varying envelopes.  We thus suppose that the atomic field is
characterized by a central wave vector $\bk_0$ and try an expansion of
the form
\begin{equation}
\label{eq:slow-expand}
\psi(\bx) = \sum_n \fnbk(\bx,t) \ket{n\bk_0} e^{-iE_{n\bk_0}t/\hbar}.
\end{equation}
where the envelope functions $\fnbk(\bx,t)$ are assumed to vary slowly
on the scale of the lattice period.  This is just the case for atomic
BECs or an atom laser, as the momentum distribution is always
distributed narrowly about some central wave vector. We defer the
question of the choice of $\bk_0$ and the calculation of the initial
envelopes $\fnbk(\bx,t=0)$ for a moment and concentrate on the
dynamics.

Substituting Eq.~\eqref{eq:slow-expand} into the governing
Eq.~\eqref{eq:gpe} gives a complicated equation involving sums over
all the envelope functions and still involving the rapidly oscillating
periodic potential.  To obtain a simpler description, one insists that
the variation of the $\fnbk$ with $\bx$ occur on slower scales than
that of the Bloch functions $\ket{n \bk_0}$.  Such a condition may be
applied using the effective mass formalism of solid state
theory~\cite{callaway}, or equivalently, using ``multiple scales''
techniques as has been demonstrated in the context of nonlinear optics
in a series of papers by de~Sterke, Sipe and
co-workers~\cite{des94,des96}.  Essentially, one obtains an equation
for each envelope by dotting from the left with each Bloch function
$\bra{n \bk_0}$.

In fact, if more than two envelopes are included the resulting
equations remain quite complicated. For a start, all the bands are
coupled by the nonlinearity, but there are also linear couplings that
occur between bands with a small energy separation~\cite{des96}.
However, for many problems of interest, it may be sufficient to
consider just one or two bands.  For example, the stationary ground
state solution must attain the lowest energy possible. Providing the
nonlinearity is not too large, the dominant contribution to the
solution is then provided by the lowest band at $\bk_0=\mathbf{0}$.
If the nonlinearity is very large, coupling to the higher bands can
not be ignored and either more bands must be included or a different
approach adopted. We return to this point in
section~\ref{sec:strong-nonlin}.  A one band approach can also be
acceptable for time-dependent problems.  A natural experiment might be
to form a condensate in the stationary ground state of the lattice and
then apply a weak momentum kick via a momentary change in the
potential. In this case, we would expect the solution to be well
represented by an expansion on the lowest band but concentrated around
a nonzero value $\bk_0$.  If the system is tuned close to a band gap,
however, there is strong coupling between the Bloch modes above and
below the gap, and a two-mode approach is necessary~\cite{des94}.  We
consider this situation in section~\ref{sec:two-modes}.

\subsubsection{Single band approach}
In calculating solutions we shall concentrate on the stationary ground
state of the BEC in the finite lattice, and hence a single envelope
function is sufficient. We emphasize that for the envelope function
approach to be valid, the nonlinearity must be considered a
perturbation. We discuss the regime of applicability imposed by this
constraint when we examine results in section~\ref{sec:weak-nonlin}.

By applying the effective mass or multiple scales
methods~\cite{callaway,des96}, we simply obtained the reduced GPE for
the envelope function $\fnbk$,
\begin{equation}
\label{eq:dynamics1}
i\hbar \left( \parpar{\fnbk}{t} + \bv_g \cdot \bdel \fnbk \right)
= -\frac{\hbar^2}{2 m^{*\mu\nu}} \frac{\partial^2}{\partial x_\mu
    \partial x_\nu}\fnbk + V(\bx)\fnbk + \Gamma^*|\fnbk|^2\fnbk,
\end{equation}
where
\begin{mathletters}
\label{eq:blochprops}
\begin{eqnarray}
\label{eq:blochprops1}
\bv_g &=& \bv_{nn}(\bk_0), \\
\label{eq:blochprops2}
\frac{1}{m^{*\mu\nu}}  &=& \frac{1}{m}+
     \sum_{p\ne n} \frac{v_{pn}^{\mu}(\bk_0)v_{np}^{\nu}(\bk_0) +
    v_{pn}^{\nu}(\bk_0)v_{np}^{\mu}(\bk_0) }{
    (E_{n\bk_0}-E_{p\bk_0})}, \\
\label{eq:blochprops3}
\Gamma^* &=& \frac{\Gamma}{\Omega}\int_{\text{cell}} d^3\bx \, |u_{n\bk_0}(\bx)|^4,
\end{eqnarray}
\end{mathletters}
and the velocity matrix elements are defined as
\begin{equation}
\label{eq:velocity}
        \bv_{ij}(\bk_0) =
        \frac{\fullexpect{i\bk_0}{\hatbp}{j\bk_0}}{m}
= \frac{1}{m}\left [ \hbar \bk_0 \delta_{ij}+\int_{\text{cell}} d^3\bx 
\, u_{i\bk_0}(\bx)^* \hatbp u_{j\bk_0}(\bx) \right ].
\end{equation}
Equation~\eqref{eq:dynamics1} is valid provided that the maximum
nonlinear ``detuning'' $\Gamma^*|\fnbk|^2$ at the point 
where $|\fnbk|$ attains its
largest value is small compared to the energy separation to the
nearest adjacent band at wave vector $\bk_0$.  If this is not
the case we must use a two band model~\cite{des94,des96} as discussed in 
section~\ref{sec:two-modes}. We now have a
significant advance in that the periodic potential does not appear
explicitly in Eq.~\eqref{eq:dynamics1}.  Its action is effectively
represented through the three parameters in Eqs.~\eqref{eq:blochprops}
which we now discuss.  The envelope $\fnbk$ moves with a group
velocity $\bv_g$ which is simply the gradient of the band $n$ at
wavevector $\bk_0$.  Note that at particular points of symmetry the
group velocity vanishes, notably at the band edges, and of course at
the bottom of the lowest band for $n=0$, $\bk=\mathbf{0}$.  The bare
atomic mass $m$ has been replaced by the effective mass tensor $m^*$
which is determined by the interaction of the other bands with the
band of interest~[Eq.~\eqref{eq:blochprops2}].  The term involving the
effective mass in Eq.~\eqref{eq:dynamics1} acts as a source of
dispersion on the atomic field.  Taken together the group velocity and
effective mass constitute a quadratic expansion of the band $n$ around
the point $\bk_0$.  Finally, the parameter $\Gamma^*$ describes the
change in the effective nonlinearity due to the shape of the Bloch
function. With a weak potential, for which $u_{n\bk}$ is almost a
constant function, $\Gamma^* \approx \Gamma$.  However with a strong
lattice, for which the Bloch functions become peaked within or between
the wells, the magnitude of the effective nonlinearity increases
significantly.

The wavevector $\bk_0$ is chosen in the first Brillioun zone and may
be known from the nature of the problem---say an atom laser pulse of
known momentum incident on the optical lattice---or might be
calculated using an appropriate definition such as
\begin{equation}
\bk_0 = \int d\bk^3 \, |\tilde{\psi}(\bk,t=0)|^2 \bK(\bk)
\end{equation}
where $\tilde{\psi}(\bk)$ is the Fourier transform of $\psi(\bx)$ and
$\bK(\bk)$ is the vector in the first Brillioun zone equivalent to
$\bk$.  If a stationary state is sought we of course have $\bk_0=0$.

\subsubsection{Soliton solutions}
Equation~\eqref{eq:dynamics1} is a nonlinear Schr\"{o}dinger equation
(NLSE) describing the evolution of the envelope field as it propagates
through the lattice under the influence of the atomic interactions and
the external potential $V(\bx)$. Note that for one-dimensional
geometries in the limit of vanishing $V(x)$ we can obtain soliton
solutions for the envelope. It must be emphasized that it is the
envelope function that has soliton solutions, hence these are not the
same objects as the standard atomic solitons found when the field
operator or wave function itself satisfies a
NLSE~\cite{len93,zha94,mor97,dyr97}. The envelope soliton propagates
at a velocity determined by the local gradient of the band structure
and with a width determined by a balance between the nonlinearity and
the effective mass, rather than the bare mass.  Such ``Bragg grating
solitons'' have been observed in recent years in optical fiber
gratings at extremely high laser powers~\cite{egg96,bro97b}.  In the
present case, the nonlinearity is much stronger and BEC grating
solitons should be easier to generate.  For example, our calculations
suggest that for sodium atoms in a weak lattice generated by a diode
laser of wavelength 985~nm (see section~\ref{sec:weak-nonlin}), a
condensate of 500 atoms could produce a Bragg soliton extending over
30 periods of the lattice.

For time-dependent problems, the term in $V(\bx)$ in
Eq.~\eqref{eq:dynamics1} describes the response of the soliton to slow
changes in the non-periodic potential. For example a broad laser beam
incident from the side could be used to produce a local ``hill'' in
the optical lattice.  Gradients in this potential of course generate
spatial oscillations in the field, which is to say, they induce wave
vector shifts. For modest wave vector shifts, Eq.~\eqref{eq:dynamics1}
correctly describes the resulting changes in velocity and shape of the
soliton. However, if the shift becomes too large, the quadratic
expansion of the Bloch band implicit in the effective mass
approximation breaks down, and there must be a dynamic reallocation of
the coefficients $\bv_g$, $m^*$ and $\Gamma^*$ corresponding to the
new point on the band~\cite{mjs94}.  Alternatively, one could extend
the effective mass approach by including higher spatial derivatives in
Eq.~\eqref{eq:dynamics1}.

\section{Stationary solutions} \label{sec:Station-solutions}
While the possibility of Bragg solitons in BEC's is interesting, it is
the properties of the ground state solution of a BEC in a lattice
which are initially likely to be of most importance for experiments,
and it is on these that we focus.  We shall concentrate on the
one-dimensional geometry which is perhaps the most pertinent
experimentally.  The harmonic trap potential $V(\bx)$ is assumed to be
tightly confining in the directions transverse to the lattice
potential and weakly confining along the lattice, such that the trap
angular frequencies in the perpendicular ($\omega_r$) and axial
($\omega_z$) direction satisfy $\omega_r \gg \omega_z$.
Eq.~\eqref{eq:gpe} then reduces to
\begin{equation}
  \label{eq:gpestat}
  \mu \psi(z) = \left[-\frac{\hbar^2}{2m}\frac{d^2}{dz^2}
        -\kappa \cos(k_z z) +\frac{1}{2} m \omega_z^2 z^2 + 
        N\tilGam |\psi(z)|^2 \right] \psi(z),
\end{equation}
where $\tilGam=\Gamma/A_{\text{eff}}=2\hbar \omega_r a$ is the
effective nonlinearity taking account of the transverse mode profile.
There are two natural choices of dimensionless variables: the usual
set of harmonic oscillator units and a set in which the period of the
lattice is unity. We shall find it useful to use both at different
times.  The sets are introduced by choosing appropriate energy and
distance scales.  For the oscillator case, these are $\epsilon_o=
\hbar \omega_z$ and $Z_o = \sqrt{\hbar/(m\omega_z)}$ respectively.
For the lattice case we instead have $\epsilon_l= \hbar^2/(m d^2)$ and
$Z_l = d$.  We will use an overbar to denote quantities in oscillator
units and a caret for quantities in lattice units.  Thus in oscillator
units we have $\zeta = z/Z_o$, $f(\zeta)=\sqrt{Z_o}\psi(z)$, $\barLam
= \lambda Z_o$, $\barkap=\kappa /\epsilon_o$, $\barmu_o=\mu
/\epsilon_o$, $\bar{\Omega}_0 = \hbar \omega_z/\epsilon_o =1$ and
$\bar{C}=N \tilGam /(\epsilon_o Z_o)$.  Similarly for the lattice
case, we have $\xi = z/Z_l$, $g(\xi)=\sqrt{Z_l}\psi(z)$ and $\carLam =
\lambda Z_l=2\pi$, while the other parameters are defined in the
obvious fashion, replacing the subscripts $o$ by $l$ and the overbars
by carets.

We have included the number of atoms $N$ in the nonlinear constants
$\barC$ and $\carC$ and so the wave functions are normalized to unity.
We now turn to finding the ground state solutions of
Eq.~\eqref{eq:gpestat}, identifying a number of regimes depending on
the relative strength of the interactions and the lattice.

\subsection{Weak nonlinearity} \label{sec:weak-nonlin}
The regime of a weak nonlinearity is handled well by the effective
mass envelope function approach introduced in
section~\ref{sec:method}.  In this case the nonlinearity and the
harmonic trap both act as perturbations to the dominant lattice
potential and it is convenient to use lattice units. Thus we seek
solutions to
\begin{equation}
  \label{eq:latstat}
  \carmu g(\xi) = \left[-\frac{1}{2}\frac{d^2}{d\xi^2}
        -\carkap \cos(2\pi \xi) +\frac{1}{2} \carOmega^2 \xi^2 + 
        \carC|g(\xi)|^2 \right] g(\xi).
\end{equation}
The wave function is written $g(\xi) = a(\xi) \phi_0(\xi)$ where
$\phi_0(\xi)$ is the lowest energy Bloch function ($k=0$ in the lowest
band) for a lattice of depth $\carkap$ and period 1, and has
eigenvalue $\carE_{00}(\carkap)=E_{n=0,k=0}/\epsilon_l$. In the
lattice units, Eq.~\eqref{eq:dynamics1} becomes
\begin{equation}
        \label{eq:effmass}
  \hat{E} a(\xi) = \left[-\frac{1}{2\carm^*}\frac{d^2}{d\xi^2}
        +\frac{1}{2} \carOmega^2 \xi^2 + C^*|a(\xi)|^2 \right] a(\xi).
\end{equation}
As the lattice period in Eq.~\eqref{eq:gpestat} is unity, the shape of
the Bloch function $\phi_0(\xi) $ is specified by the single parameter
$\carkap$, as are the values of the nonlinearity
$\carC^*=\gamma(\carkap) \carC$, with
\begin{equation}
\label{eq:effnl}
\gamma(\carkap)= \int_0^1 d\xi \, |\phi_0(\xi)|^4,
\end{equation}
and the (now scalar) effective mass given by
\begin{equation}
  \label{eq:scalarmass}
\frac{1}{\carm^*}=1+2 \sum_{p>0} \frac{|\fullexpect{\phi_p(\xi)}{-i
    \partial/\partial \xi}{\phi_0(\xi)}|^2}{\carE_{00}-\carE_{p0}},  
\end{equation}
where the Bloch functions $\phi_p(\xi)$ and energies $\carE_{p0}$ are
of course normalized in lattice units.  If we did not use lattice
units, we would be forced to consider the dependence of these
quantities on the lattice period as well.  The effective quantities
for the cosine potential are plotted in Fig.~\ref{fig:effective} as a
function of $\carkap$.  Note that in the limit of a vanishing lattice
potential, $\carm^*=1$ and $\gamma(\carkap)=1$. For large $\carkap$,
the Bloch function tends to a sequence of harmonic oscillator ground
state wave functions and $\gamma(\carkap)\rightarrow \carkap^{1/4}$.

The importance of Eq.~\eqref{eq:effmass} is that the problem including
the lattice has now been reduced to the usual GPE for a condensate in
just a harmonic trap with no lattice. All the physics involving the
lattice is contained in the effective parameters and we are left with
an equation whose properties are well understood.
Equation~\eqref{eq:effmass} may naturally be solved in identical
fashion to the standard GPE---numerically in general, or using
Gaussian or Thomas-Fermi approximations (TFA) in appropriate limits.
In particular, the TFA solution has the form
\begin{equation}
  \label{eq:actualEMTFA}
  |a(\xi)|^2=\max\left\{ \frac{\carE-\carOmega \xi^2/2}{\carC^*},0\right\}.
\end{equation}
Note that as the effective mass grows rapidly with $\carkap$ (see
Fig~\ref{fig:effective}), the TFA to Eq.~\eqref{eq:effmass} is valid
for quite modest values of $\carkap$.  It must be emphasized though,
that physically, this ``effective mass TFA'' (EMTFA) is a rather
different approximation to the standard TFA which applies when the
total kinetic energy is negligible.  Here it is only the kinetic
energy associated with the \emph{envelope function} which may be
considered small. Indeed there is considerable kinetic energy
associated with the oscillation in the Bloch function but this of
course has been removed from Eq.~\eqref{eq:effmass}. It must also be
realized that even if Eq.~\eqref{eq:effmass} is solved numerically,
this represents a significant gain over a direct numerical solution of
the complete GPE with the lattice potential included.  From a
practical point of view, the numerical routine is much faster as it
requires a small fraction of the grid points required to simulate the
whole lattice. This advantage is most obvious when one considers
time-dependent problems using Eq.~\eqref{eq:dynamics1} rather than
Eq.~\eqref{eq:gpe}. Less prosaicly, the separation of scales allows a
clearer physical insight into the expected general shape of the ground
state condensate, with the precise details of the rapid oscillation
subsumed into a few parameters, that need only be calculated once.

Let us now consider the regime of validity for the effective mass
approach as a whole and the EMTFA. To justify an expansion in terms of
the linear Bloch modes of the lattice, the effect of the nonlinearity
in Eq.~\eqref{eq:latstat}
must be small compared to that of the lattice potential, \emph{on the 
scale of a single period}. This is the sense in which the nonlinearity
is to be regarded as a perturbation.
Thus if $\rho=g(\xi)^2$ is the density of
the envelope function, we require
\begin{equation}
\label{eq:EM-cond}
\carC \rho \ll \carkap.
\end{equation}
Assuming that this is so, it may be checked that the EMTFA to
Eq.~\eqref{eq:effmass} is valid if
\begin{equation}
\label{eq:EMTFA-cond}
\carC^* \gg 2 \sqrt{\carOmega}/(\carm^*)^{3/4},
\end{equation}
Using the EMTFA expression for the peak envelope density at the
origin, Eq.~\eqref{eq:EM-cond} may then be expressed as
\begin{equation}
\label{eq:EMTFA-cond2}
\carOmega \carC \ll \frac{2}{3} \sqrt{\gamma(\carkap)(2\carkap)^3}.
\end{equation}
Note that due to its weak dependence on $\carkap$, the dependence on
$\gamma(\carkap)$ may usually be ignored.  If the EMTFA is not valid,
we can obtain a different bound on $\carC$ by noting that the peak
density of the envelope will certainly be lower than for the
associated linear system for which $a(\xi)=1/\sqrt{r_0
  \sqrt{\pi}}\exp(-\xi^2/(2r_0^2))$, with $r_0=1/\sqrt{\carOmega
  \carm^*}$.  Thus we obtain the condition
\begin{equation}
\label{eq:lin-cond}
\carC \ll \carkap \sqrt{\frac{\carOmega \carm^*}{\pi}} .
\end{equation}
But as Eq.~\eqref{eq:EMTFA-cond} is not true, Eq.~\eqref{eq:lin-cond}
is satisfied provided only that $\carOmega \ll \sqrt{\pi}\carkap$.
These conditions are summarized in Table~\ref{tab:conditions} along
with the chemical potential which is given by
$\carmu=\carE_{00}(\carkap)+\carE \rightarrow
\carE_{00}+(3\carC^*\carOmega/2)^{2/3}/2$, where the second relation
holds when the EMTFA is valid.  To demonstrate that these conditions
could be satisfied by current experiments we consider $\mbox{Na}^{23}$
atoms with a lattice produced by counter propagating beams from a
diode laser at 985~nm~\cite{sta98} so that $d = 0.49$~$\mu$m.  
Taking a detuning of $\Delta=10^{10}\mbox{
  s}^-1$ and a maximum Rabi frequency of $\Omega_e=0.01 \Delta$ we
obtain a maximum lattice depth of
$\carkap_{\text{max}}=\kappa_{\text{max}}/\epsilon_l=\hbar \Omega_e^2/(\epsilon_l 
\Delta)\approx 90$.  Assuming $\omega_r/(2\pi . 50) =\omega_z/(2\pi) =
20$~Hz and a scattering length $a=5$~nm gives 
$\carOmega=0.01$ and $\carC=0.013N$.  Choosing
$\carkap=90$, we could satisfy Eq.~\eqref{eq:EMTFA-cond2} with
condensates of up to a few million atoms.

We show two examples of the envelope approach in Fig.~\ref{fig:weakC},
comparing the condensate density as a function of position according
to the various models discussed above.  For both plots the
nonlinearity $\carC=1$ and trap frequency $\carOmega=0.05$.
Fig.~\ref{fig:weakC}a shows a case for a weak lattice with
$\carkap=10$, while in Fig.~\ref{fig:weakC}b, the lattice strength is
$\carkap=100$.  The results of exact numerical calculations using
Eq.~\eqref{eq:latstat} are shown by the solid line in each figure. For
clarity, we show only the positive $\xi$ part of the solution; the
negative part is an exact mirror of this.  The overall extent of the
wave function is similar in the weak and strong lattices with the weak
lattice case appearing to have a longer tail in the condensate
density. In the stronger lattice, the oscillations are clearly more
pronounced with the density confined more closely to the center of
each well.  

Turning to the accuracy of our approximate descriptions, for the weak
case in Fig.~\ref{fig:weakC}a, the effective mass description of
Eq.~\eqref{eq:effmass} is shown with a dotted line which is barely
distinguishable from the exact result.  The dashed line indicates the
EMTFA solution which provides reasonable agreement with the exact
solution near the center of the trap but fails towards the edge where
the neglect of the envelope's kinetic energy with respect to the
nonlinear part is invalid.  For the strong lattice
(Fig.~\ref{fig:weakC}b), the effective mass and EMTFA results are now
perfectly superposed and are both indicated by the dashed line which
has excellent agreement with the exact case for all but the last
two peaks of the distribution.  These results concur with the validity
regime calculations above: due to the difference in effective masses,
Eq.~\eqref{eq:EMTFA-cond} predicts that the EMTFA should be valid for
the strong lattice but not for the weak.

An additional dot-dashed line in each figure indicates the exact
solution of Eq.~\eqref{eq:gpestat} when the nonlinearity vanishes
($\carC=0$). Comparing the two linear solutions shows how in the
absence of a nonlinearity, the condensate shrinks with increasing
lattice strength due to the increase in effective mass.  In both
figures, the exact solution is clearly broader than the linear result,
despite the repulsive effect of the kinetic energy being reduced
through the effective mass. This reduction is of course outweighed by
the dominant repulsive nature of the nonlinearity. In spite of this,
the \emph{linear} Bloch function is perfectly adequate to represent
the rapid oscillation---the nonlinearity is not strong enough to
distort the wave function on the scale of a single period.

Finally we consider the relative density distributions for the two
exact solutions. Increasing the lattice strength produces two
competing effects---the increase effective mass reduces the repulsion
due to the kinetic energy, while the increase in effective
nonlinearity increases the interatomic repulsion. For large enough
effective mass, however, the kinetic energy plays essentially no role
and any further increase in the mass makes no difference, whereas the
effective nonlinearity and therefore the nonlinearly-induced repulsion
grows slowly but without bound.  Hence, we should expect that the
condensate spreads out with increasing lattice strength.  In
Fig.~\ref{fig:density}, we plot the cumulative density
$\int_{-\xi}^\xi d\xi' \, |g(\xi')|^2 $ as a function of $\xi$ for
several values of the lattice strength $\carkap$. From this plot it is
clear that in spite of the weakening kinetic energy, the condensate
slowly expands as the lattice strength ranges over
$\carkap=0$~(dot-dashed line), 10~(solid), 50~(dotted) and
100~(dashed), due to the increase in the effective nonlinearity [see
Eq.~\eqref{eq:effnl}]. The second dot-dashed line indicates the
cumulative density according to the EMTFA envelope for $\carkap=100$.
Note that the exact solution in the weak lattice appears to have a
longer tail than for the strong lattice (solid lines in
Figs.~\ref{fig:weakC}).  Figure~\ref{fig:density}
shows that the structure of the tail is not indicative of the global
density distribution.

\subsection{Strong nonlinearity} \label{sec:strong-nonlin}
As the nonlinearity increases or the trap tightens, the assumption
that the nonlinearity plays no role over a single period
[Eq.~\eqref{eq:EM-cond}] breaks down.  While one may use additional
envelope functions each with their own Schr\"{o}dinger equation, the
utility of such an approach decreases quickly.  However, if the
nonlinearity is sufficiently large, we may abandon the Bloch function
approach altogether, instead using the standard Thomas-Fermi
approximation (TFA) by neglecting the kinetic energy term in
Eq.~\eqref{eq:gpestat}.  S{\o}renberg and M{\o}lmer~\cite{sor98} have
used this approximation for the \emph{infinite uniform} lattice. 
Here, we examine the more complicated system including
the trap potential in the Thomas-Fermi regime.
In this regime, the lattice is effectively a
perturbation to the dominant harmonic trap, and it is natural to
invoke oscillator units, so that Eq.~\eqref{eq:gpestat} becomes
\begin{equation}
  \label{eq:oscstat}
  \barmu f(\zeta) = \left[-\frac{1}{2}\frac{d^2}{d\zeta^2}
        -\barkap \cos(\barLam \zeta) +\frac{1}{2} \zeta^2 + 
        \barC|f(\zeta)|^2 \right] f(\zeta).
\end{equation}
\begin{equation}
\label{eq:strongTFA}
|f(\zeta)|^2 = \mathrm{max}\left \{ \frac{\barmu-\frac{1}{2}\zeta^2 +
\barkap \cos(\barLam \zeta)}{\barC}, 0\right \}.
\end{equation}
In estimating the boundary point $\zeta_0$ 
of the condensate we may neglect the effect
of the lattice, and take the usual value $\zeta_0=\sqrt{2 \barmu}$.
This is consistent with the nature of the approximation which is poor
at the boundary anyway.  Moreover, to the same degree of approximation
the dependence of the chemical potential on atom number is unchanged
from the lattice free result $\barmu=(3\barC/2)^{2/3}/2$.  

The regime of applicability for the TFA is analyzed as follows.  We
must certainly have $\barmu > \barkap$ which prevents the solution
from attaining zero density regions in every lattice well.  The
standard TFA condition that the total kinetic energy be negligible
compared to the total potential and nonlinear energies is given by
$\barC \gg 1$ as for the lattice-free case.  However, if the
oscillatory part of the solution is to be accurately represented by
the factor $\barkap \cos(\barLam \zeta)/\barC$ in
Eq.~\eqref{eq:strongTFA} we
must also consider the relative energies associated with \emph{just the
oscillatory part}.  An approximate condition may be found by 
evaluating these energies over a single lattice period at the center
of the trap and insisting as usual that the resulting kinetic energy
be small compared to the potential and nonlinear energies.
Assuming $\barkap\ll \barmu$ we eventually arrive at the additional condition
\begin{equation}
\label{eq:tfcond1}
\barmu \gg \barLam^2/4 ,
\end{equation}
or
\begin{equation}
\label{eq:tfcond2}
\barC \gg 0.24 \barLam^3.
\end{equation}
Note that in the lattice units, this has the slightly odd form $\carmu
\gg \pi^2$. We see below that this new condition is genuinely necessary.

We show examples in Fig.~\ref{fig:thomas} with exact numerical
solutions to Eq.~\eqref{eq:oscstat} in solid lines, and the TFA in
dotted lines. In all three plots we have $\barC=3000$. In
Fig.~\ref{fig:thomas}a the lattice strength $\barkap=10$, and
$\barLam=2\pi$. The exact chemical potential is $\barmu=136.2$.  The
requirements for the TFA are thus satisfied and there is good
agreement between the exact and approximate solutions except near the
boundary of the condensate.  Figure~\ref{fig:thomas}b shows the
solutions where the lattice strength is increased to $\barkap=50$, for
which the exact chemical potential is $\barmu=133.8$.  The region of
disagreement near the boundary is now of course larger, but there is
still good agreement near the center of the condensate despite
$\barkap$ approaching 50~\% of the chemical potential.  In
Fig.~\ref{fig:thomas}c, we have $\barkap=50$ with a reduced lattice
period such that $\barLam=5\pi$. The chemical potential is
$\barmu=134$.  As the ratio $\barmu/(\barLam^2/4)=0.54$,
Eq.~\eqref{eq:tfcond1} indicates the TFA is invalid, in accord with
the poor agreement  between the TFA and exact curves.  We emphasize
that if the lattice were not present, the TFA would certainly be valid
in this case because $\barC \gg 1$. Hence the condition given by
Eq.~\eqref{eq:tfcond1} is indeed important. The dashed lines in
Fig.~\ref{fig:thomas}c indicate the extent of oscillations in the
solution predicted by the single band Bloch function description used
in the previous section for weak nonlinearity.  This is clearly
unsuccessful in this regime as we would expect. Comparing it to the
 exact solution
indicates that the repulsive nature of the nonlinearity tends to
reduce the amplitude of the oscillations as compared to the linear
regime.  Note that despite the range in shapes of the three exact
solutions in Figs.~\ref{fig:thomas}a-c, the chemical potential varies
by only a few per cent from the lattice free value $\barmu=136.2$.
The parameter values used here would be easily obtained
experimentally. Using the same trap frequencies $\omega_r/(2\pi . 50)
=\omega_z/(2\pi) = 20$~Hz, we find a maximum lattice strength of
$\barkap_{max}\approx 8000$ and an effective nonlinearity of
$\barC=0.1 N$. Hence, a condensate of about $N=30000$ atoms would
correspond to the predictions in Figs.~\ref{fig:thomas}. Note that
when expressed in equivalent units, the
lattice strength here is around 1~\% of that in
section~\ref{sec:weak-nonlin}, which explains why condensates of a
million atoms could be considered to be weakly nonlinear in that
section.

It is worth noting that in the TFA regime, an envelope function
approach is not appropriate, because the amplitude of the oscillations
in the condensate density due to the lattice are of the same size
across the whole condensate.  In contrast, we expect an envelope
function approach to be appropriate if the rapid oscillations in the
density scale linearly with the envelope density.  This difference is
clearly apparent if the exact solutions in Figs.~\ref{fig:weakC}
and~\ref{fig:thomas} are compared.

\section{Discussion: atomic fields near a band edge} \label{sec:two-modes}
Finally, we return to the propagation problem and briefly discuss the
rather different physics involved for condensates with a center wave
vector tuned close to a band gap.  In this case, unless the gap is
very large and the field is tuned close to one edge, then neither of
the two bands surrounding the gap can be regarded as remote and the
field should be expanded using a Bloch function from each band.  As
our primary interest has been in the ground state stationary solution
which is always far from any gap, we shall not explore this case in
detail. For the sake of completeness, however, we state the dynamical
equations that correctly describe such a situation using a method of
de~Sterke~\emph{et al.}~\cite{des96}.

We label the two bands $u$ (upper) and $l$ (lower) and expand around a
frequency $\omega_0= (\omega_{u\bk}+\omega_{l\bk})/2$:
\begin{equation}
\label{eq:ansatz2}
\psi(\bx,t) =\left[\fubk(\bx,t) \ket{u\bk} + \flbk(\bx,t) \ket{l\bk}\right]
e^{-i \omega_0 t}.
\end{equation}
Again one insists that both $\fubk$ and $\flbk$ are slowly-varying
functions and using the effective mass or multiple scales techniques
obtains two coupled nonlinear equations for $\fubk$ and $\flbk$. In
coupled mode theory of periodic structures it is often found
convenient to work with envelopes modulating forward and backward
going amplitudes as opposed to Bloch functions as these envelopes
correspond closely to the field incoming and outgoing from the
periodic structure. It can be shown that
the amplitudes $G_\pm=(\flbk \mp i \fubk )/2$ correspond to such
amplitudes.  In the limit of a weak potential where the Bloch
functions at the band edge are cosine functions, the amplitudes
$G_\pm$ simply modulate forward and backward-going plane waves.  In
terms of these variables the final dynamical equations
are~\cite{des96}
\begin{mathletters}
\label{eq:dynamics2}
\begin{eqnarray}
&i&\parpar{G_+}{t} + \bar{\bv}_g \cdot \bdel G_+ + \sigma G_- +V(\bx) G_+
+ \Gamma_0 (|G_+|^2 + 2|G_-|^2) G_+ \nonumber \\
~ &~&\hspace{1cm} + \Gamma_1 (|G_+|^2 + |G_-|^2) G_- + \Gamma_1(G_+
G_-^*+G_+^* G_-) G_+
+ \Gamma_2 G_-^2 G_+^* =0, \\
&i& \parpar{G_-}{t} - \bar{\bv}_g \cdot \bdel G_- + \sigma G_+ +V(\bx) G_-
+ \Gamma_0 (|G_-|^2 + 2|G_+|^2) G_-  \nonumber \\
~ &~&  \hspace{1cm} + \Gamma_1 (|G_+|^2 + |G_-|^2) G_+ + \Gamma_1(G_+
G_-^*+G_+^* G_-) G_- + \Gamma_2 G_+^2 G_-^* =0.
\end{eqnarray}
\end{mathletters}
The linear coupling parameter $\sigma=\Delta/2$, the group velocity
$\bar{\bv}_g=|\fullexpect{u\bk}{\hatbp}{l\bk}|$, while the nonlinear
coefficients are found from overlaps of the (real) Bloch functions:
\begin{align}
  \Gamma_0 &= \frac{\alpha_{llll}+2 \alpha_{uull}+\alpha_{uuuu}}{2} , \\
  \Gamma_1 &= \frac{\alpha_{llll}-\alpha_{uuuu}}{2}, \\
  \Gamma_2 &= \frac{\alpha_{llll}-6 \alpha_{uull}+\alpha_{uuuu}}{2},
\end{align}
where
\begin{equation}
\alpha_{ijkl} = \frac{\Gamma}{\Omega}\int_{\text{cell}} d^3 \bx 
        \phi_i(\bx)\phi_j(\bx)\phi_k(\bx)\phi_l(\bx)
\end{equation}
for $\{i,j,k,l\}$ in $\{u,l\}$.  In these equations, we have dropped
the kinetic energy terms involving the effective mass altogether. The
coupling between the bands through the parameter $\sigma$ itself
produces a dispersive behavior which is much larger than the intrinsic
dispersion within each band.  In the limit of a weak potential, the
coefficients $\Gamma_1$ and $\Gamma_2$ vanish and one obtains the
well-known coupled mode equations for propagation in nonlinear
periodic systems. Such equations were recently obtained in the BEC
context by Zobay~\emph{et al.}~\cite{zob98} and used to predict the
occurrence of gap solitons in condensates with multiple internal
levels.  Their system differs from ours in that it involves a Raman
coupling between two magnetic ground states whereas we have used a
strictly scalar field interacting with an external potential.  The
fact that both systems can be reduced to identical equations is an
indication of the underlying physical consequences of combining a
nonlinearity with a periodic system---properties such as gap solitons
and switching appear ubiquitously.  The full ``deep grating''
equations~\eqref{eq:dynamics2} with $\Gamma_1, \Gamma_2 \ne 0$ (but
$V(\bx)=0$ were first derived by Salinas~\emph{et al.}~\cite{sal94} to
describe optical propagation in Bragg gratings with \emph{large} index
variation between the two media.  The additional terms in these
equations lead to predictions that depart from the shallow grating
results for index variations of 10~\% or more.  In the atomic case,
the analogous situation is when the lattice potential is deep and the
Bloch functions at the band edge depart from simple plane waves.  The
exposure of condensates to such deep lattices is certainly a realistic
goal in the laboratory and the use of the deep grating equations may
thus be important in modeling propagation in lattices in the future.
These arguments hold equally for the Raman geometry used by
Zobay~\emph{et al.}

We should remark that the use of a two band model as described here is
not always possible. In the case of extremely deep potentials the band
gap widths may broaden so much they become comparable to the original
separations between the gaps. In that case, the existence of two bands
that are significantly closer to the center frequency than all the
others is doubtful. However, if the desired center frequency is closer
to one band edge than the other, a one band model may again become
feasible.

\section{Conclusion}
We have introduced a Bloch function analysis of a BEC trapped in a
finite optical potential. The approach describes both pulse
propagation and stationary solutions providing a single
Schr\"{o}dinger like equation for the envelope function provided the
nonlinearity is not too large. If the physical parameters lie within
the constraints we have given, the envelope function solution agrees
extremely well with the exact numerical results. For very strong
nonlinearities, the standard Thomas-Fermi approximation is the correct
description for the ground state, but its application requires a
stronger condition than in the absence of the lattice.

\acknowledgements We thank C. Martijn de Sterke and Karl-Peter Marzlin
for useful discussions. This research was supported by the Australian
Research Council.

\begin{figure}
\caption{\label{fig:effective} Effective nonlinearity $\Gamma^*$ (solid) 
  and mass $m^*$ (dotted) for a one-dimensional cosine potential of
  strength $\kappa$. }
\end{figure}

\begin{figure}

\caption{\label{fig:weakC} Wave function $\psi(\xi)$ as a function of
  $\xi$ for a) $\carkap=10$ and b) $\carkap=100$.  Other parameters
  are $\carOmega=0.05$ and $\carC=1$. Line styles indicate models as
  exact numerical (solid), effective mass (dotted), EMTFA (dashed) and
  linear (dot-dashed). In b), the effective mass and EMTFA models are
  both represented by the dashed curve. }
\end{figure}

\begin{figure}
\caption{\label{fig:density} 
  Cumulative density $\int_{-\xi}^\xi d\xi'\, |g(\xi')|^2 $ as a
  function of $\xi$ for values of the lattice depth $\carkap=10$
  (solid line), 50 (dotted) and 100 (dashed).  The dot-dashed lines
  indicate the cumulative density of the exact solution for
  $\carkap=0$, and of the EMTFA envelope for $\carkap=100$.}
\end{figure}

\begin{figure}
\caption{\label{fig:thomas} Exact wave function $f(\zeta)$ (solid line)
  and Thomas-Fermi estimate (dotted) with nonlinearity $\barC=3000$
  and a) $\barkap=10$, $\barLam=2\pi$, b) $\barkap=50$, $\barLam=2\pi$
  and c) $\barkap=50$, $\barLam =5\pi$ .  The dashed line in c)
  denotes the prediction of the envelope function approach.}
\end{figure}

\begin{table}
\begin{tabular}{lll}
Regime    & Conditions               & Chemical potential          \\
\hline
Weak: $\carC \rho\ll \carkap$ , (EMTFA)  & 
$\carC^* \gg 2 \sqrt{\carOmega}/(\carm^*)^{3/4}$,
$\carOmega \carC \ll \frac{2}{3} \sqrt{\gamma(\carkap)(2\carkap)^3}$
& $\carmu=\carE_{00}(\carkap)+(3\carC^*/2)^{2/3}/2$ \\
Weak: $\carC \rho\ll \carkap$ , (non-EMTFA)  & 
$\carC \ll \carkap \sqrt{\carOmega \carm^*/\pi} $
& $\carmu=\carE_{00}(\carkap)+ \carE$ \\
Thomas-Fermi:        & $\barmu \gg \barLam^2/4$, $\barC \gg 0.24 \barLam^3$
& $\barmu\approx(3\barC/2)^{2/3}/2$ \\  
\end{tabular}
\caption{\label{tab:conditions} Validity criteria and chemical
potential for ground state solutions in different regimes.}
\end{table}


\begin{thebibliography}{10}

\bibitem{newbecs}
M. H. Anderson \emph{et al}, Science \textbf{269},198 (1995); K. B. Davis
  \emph{et al}, Phys. Rev. Lett. \textbf{75}, 3969 (1995); C. C. Bradley
  \emph{et al}, Phys. Rev. Lett. \textbf{78}, 985 (1997).

\bibitem{sor98}
K. Berg-S{\o}rensen and K. M{\o}lmer, Phys. Rev. A {\bf 58},  1480  (1998).

\bibitem{zob98}
O. Zobay, E.~M. Wright, and P. Meystre, cond-mat/9805228  (1998).

\bibitem{jak98b}
D. Jaksch {\it et~al.}, cond-mat/9805329  (1998).

\bibitem{nistlaser}
L. Deng {\it et~al.}, ``Bose Einstein condensation of sodium atoms in a TOP
  trap,'' in \emph{International Quantum Electronics Conference}, Vol. 7, 1998
  OSA Technical Digest Series (Optical Society of America, Washington DC,
  1998).

\bibitem{win79}
H.~G. Winful, J.~H. Marburger, and E. Garmire, Appl. Phys. Lett. {\bf 35},  379
   (1979).

\bibitem{ace89}
A.~B. Aceves and S. Wabnitz, Phys. Lett. A {\bf 141},  37  (1989).

\bibitem{des94}
C.~M. de~Sterke and J.~E. Sipe,  in {\em Progress in Optics}, edited by E. Wolf
  (North-Holland, Amsterdam, 1994), Vol.~XXXIII, Chap.~Gap Solitons, pp.\
  203--260.

\bibitem{egg96}
B.~J. Eggleton {\it et~al.}, Phys. Rev. Lett. {\bf 76},  1627  (1996).

\bibitem{cohen2}
C. Cohen-Tannoudji, J. Dupont-Roc, and G. Grynberg, {\em Atom-Photon
  Interactions} (John Wiley \& Sons, New York, 1992).

\bibitem{mar98}
K.-P. Marzlin and W. Zhang, cond-mat/9810085  (1998).

\bibitem{callaway}
J. Callaway, {\em Quantum theory of the solid state} (Academic Press, New York,
  1974).

\bibitem{abramowitz}
Abramowitz and I. Stegun, {\em Handbook of Mathematical Functions} (Dover, New
  York, 1970).

\bibitem{des96}
C.~M. de~Sterke, D.~G. Salinas, and J.~E. Sipe, Phys. Rev. E {\bf 52},  1969
  (1996).

\bibitem{len93}
G. Lenz, P. Meystre, and E.~M. Wright, Phys. Rev. Lett. {\bf 71},  3271
  (1993).

\bibitem{zha94}
W. Zhang, D.~F. Walls, and B.~C. Sanders, Phys. Rev. Lett. {\bf 72},  60
  (1994).

\bibitem{mor97}
S.~A. Morgan, R.~J. Ballagh, and K. Burnett, Phys. Rev. A {\bf 55},  4338
  (1997).

\bibitem{dyr97}
S. Dyrting, W. Zhang, and B.~C. Sanders, Phys. Rev. A {\bf 56},  2051  (1997).

\bibitem{bro97b}
N.~G.~R. Broderick {\it et~al.}, Opt. Lett. {\bf 22},  1837  (1997).

\bibitem{mjs94}
M.~J. Steel and C.~M. de~Sterke, Phys. Rev. A {\bf 49},  5048  (1994).

\bibitem{sta98}
D. Stamper-Kurn {\it et~al.}, Phys. Rev. Lett. {\bf 80},  2027  (1998).

\bibitem{sal94}
D.~G. Salinas, C.~M. de~Sterke, and J.~E. Sipe, Opt. Commun. {\bf 111},  105
  (1994).

\end{thebibliography}
\end{document}